\documentclass[10pt, sigconf,anonymous=false]{acmart}

\usepackage[english]{babel}
\acmYear{2018}
\copyrightyear{2018}
\setcopyright{acmcopyright}
\acmConference{CoNEXT '18}{December 4--7, 2018}{Heraklion/Crete, Greece}
\acmPrice{TBA}
\acmDOI{TBA}
\acmISBN{TBA}

\begin{document}
\title{Adaptive packet transmission in response to anomaly detection in software defined smart meter networks}

\author{Mihnea Maris, Thomas Halpin, Dubem Ezeh, Karen Miu and Jaudelice de Oliveira}
\affiliation{
\institution{Drexel University \\ECE Dept}
\streetaddress{ECE Dept}
\city{Philadelphia} 
\state{PA} 
\postcode{19104}
}
\email{{mmm475,tjh323,dae55,km42,jco24}@drexel.edu}

\begin{abstract}
In this paper, we examine a basic smart meter network topology in mininet and address the 
issue of congestion over a commodity network, proposing an adaptive algorithm 
to cope with varying grid data delivery latencies. 
    
\end{abstract}


\maketitle

\section{Introduction}
In modern power distribution systems, smart meters are becoming increasingly common as utility companies increase the utilization of real-time Automatic Meter Readings (AMR) to better understand power usage and make informed control decisions. These readings are sent over networks to a centralized Energy Management System (EMS) which will do any necessary post-processing and deliver results to operators \cite{eipr}. 

An EMS might be subject to receiving and processing measurements from a large number of meters, and these meters can have varying forms of data. These measurements can be important for state estimation and are time sensitive, so it is important that the end-to-end delay and number of packets lost are minimized. In this paper, an adaptive mechanism is proposed to vary packet size in proportion with network congestion, to shorten the end to end-to-end delay and decrease the amount of packets lost.


\section{Related Work}

In \cite{AliM}, the issue of network autonomy was addressed and the author proposed a combination of machine learning and Software-Defined Networking (SDN) concepts as a way of building the network-congestion-detection component of a larger Autonomous Network Management (ANM). Three different Decision Tree algorithms C4l.5, AdaBoostM1 and Bagging) were considered in the thesis, ultimately settling for the Bagging algorithm due to a better performance over the other two in terms of accuracy metrics. 

In \cite{Salsify}, a new architecture was proposed for real-time video, combining an improved video codec and a network transport protocol to provide a more robust video streaming experience in the face of changing network conditions. Salsify optimizes the quality of each frame being streamed to match the most recent estimate of the network's capacity. 

Our proposed adaptive algorithm is inspired on Salsify. The machine learning algorithms we are using for anomaly detection have been previously reviewed for the same purpose in \cite{Tarem} and \cite{Callegari}.


\section{System Design}
In order to run experiments, we created a simple mininet topology to simulate an existing smart meter communication network. The transmitted data was generated using measurements that mimic those coming from smart meters, for accuracy. This section will detail the design of the experiments we used to gather data for congestion control.

\begin{figure}[tp]
\centering
\includegraphics[width=90mm,scale=0.8]{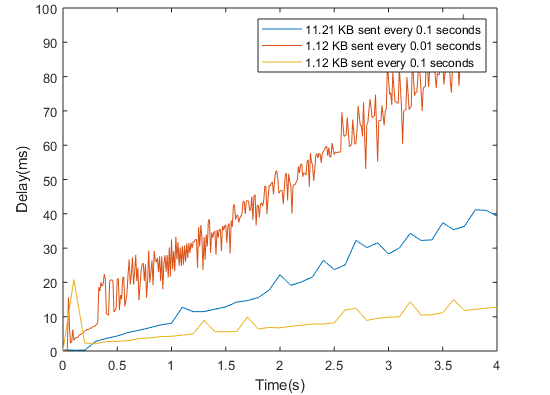}
\caption{Per packet delay comparison of packets sent at the same rate, with different packet sizes}
\label{tip}
\end{figure}

\subsection{Data collection}
In the power laboratory at Drexel University, a 3-Phase 208 Line to Line PECO source was used to power three Chroma 63800 programmable loads. These loads were connected to a Yokogawa WT1800 Power Analyzer, which recorded voltage, current, and power measurements, along with voltage and current waveforms. This data was then sent to a computer running LabVIEW software through a GPIB connection. These measurements were finally saved locally on the computer for use in later network testing. 

\subsection{Network topology}


A simple wired network topology was created in Mininet, comprising 8 host systems - one or more hosts transmitting power-related data to a receiving host while the others generated cross-traffic to congest the shared link. A single OpenFlow controller operating at the control plane is used to coordinate the activities of two Open vSwitches operating at the data plane. The switches were interconnected by means of a 100Mbps link with the same bandwidth used throughout the topology to link hosts to network devices.

The sending hosts (emulating smart meters) utilized the TCP protocol to transmit data to the receiving host; TCP was used in this work to provide guaranteed, reliable and in-order delivery of data since it has its innate congestion control algorithms and better packet loss handling. The TCP congestion control algorithm used throughout this paper is the TCP CUBIC.

\subsection{Simulation experiments}
Two main factors influence the transmission frequency of packets through the network: the size of the packet and the frequency the packets are sent at. While under normal circumstances, the packets are very small, sometimes waveform data is necessary which can easily  require thousands of data points to be transmitted per packet and cause congestion in the network. Similar to the technique used in \cite{Salsify}, we will be leveraging the resolution of the waveform to adapt the packet size to the network capacity at any given time. 

The measured data from the laboratory was replicated and modified to simulate different transmission rates and resolutions. This will show the effect the size of the packets and the transmission frequency has on the delay. Measurements were taken with and without cross traffic for reference.

\section{Simulation Results}
Multiple simulation were run for a wide range of transmission frequencies and packet sizes. To emphasize the results, we chose to focus only on 2 representative sending intervals. The 1.12 KB and 11.21 KB packet sizes correspond to data measured at a sampling period of 0.125 ms, logged and sent at 0.01 seconds and 0.1 seconds intervals respectively (higher frequency yields more data points when plotted on the same time scale). The packet sizes do not exceed the maximum transmission unit which would result in higher delay due to the packet being fragmented before sending.

Figure \ref{tip} shows a comparison between sending data at 0.1 and 0.01 intervals with their respective packet sizes. In addition, the third graph shows a lower resolution waveform sent at 0.1 seconds intervals for performance comparison. It can be observed that the slope of the graph that represents sending data at a higher frequency is larger than the ones corresponding to the lower frequency graphs. This confirms that sending data at a lower transmission rate is better during congestion conditions, since the packets sent will not add to the clogging of the system as much. 

The graph also confirms that the size of the sent message visibly affects the delay and the strain on the network under congestion conditions. We can see again that the packets transporting the lower resolution data have shorter end-to-end delay on average and a smaller slope.

\section{Future Work}
The results helped us identify that smaller packets sent at smaller rates can help transmit data more reliably through a congested network. Using this knowledge, we can use machine learning to detect anomalies and implement an algorithm that lowers the resolution of the transmitted waveform in order to keep the transmission delay in an acceptable range and reduce the risk of lost packets.

The machine algorithm used will be determined based on it's performance in our simulations. Algorithms we are considering are Kernel-based Online Anomaly Detection (KOAD), reviewed in \cite{Tarem} for detecting congestion, the Principal Component Analysis (PCA), reviewed in \cite{Callegari} and the Bagging technique used in \cite{AliM}. 

\section{Conclusion} 
We have shown how the packet size and transmission rate affect the transmission rates through a clogged network and, using the results, we are implementing a solution that will improve the performance of a smart meter communication network under similar conditions.

\bibliographystyle{ACM-Reference-Format}
\bibliography{reference}


\begin{thebibliography}{5}


\ifx \showCODEN    \undefined \def \showCODEN     #1{\unskip}     \fi
\ifx \showDOI      \undefined \def \showDOI       #1{#1}\fi
\ifx \showISBNx    \undefined \def \showISBNx     #1{\unskip}     \fi
\ifx \showISBNxiii \undefined \def \showISBNxiii  #1{\unskip}     \fi
\ifx \showISSN     \undefined \def \showISSN      #1{\unskip}     \fi
\ifx \showLCCN     \undefined \def \showLCCN      #1{\unskip}     \fi
\ifx \shownote     \undefined \def \shownote      #1{#1}          \fi
\ifx \showarticletitle \undefined \def \showarticletitle #1{#1}   \fi
\ifx \showURL      \undefined \def \showURL       {\relax}        \fi
\providecommand\bibfield[2]{#2}
\providecommand\bibinfo[2]{#2}
\providecommand\natexlab[1]{#1}
\providecommand\showeprint[2][]{arXiv:#2}

\bibitem[\protect\citeauthoryear{Ahmed, Oreshkin, and Coates}{Ahmed
  et~al\mbox{.}}{2007}]%
        {Tarem}
\bibfield{author}{\bibinfo{person}{Tarem Ahmed}, \bibinfo{person}{Boris
  Oreshkin}, {and} \bibinfo{person}{Mark Coates}.}
  \bibinfo{year}{2007}\natexlab{}.
\newblock \showarticletitle{Machine Learning Approaches to Network Anomaly
  Detection}.
\newblock \bibinfo{journal}{{\em 2nd USENIX workshop on Tackling computer
  systems problems with machine learning techniques\/}} (\bibinfo{year}{2007}),
  \bibinfo{pages}{1--6}.
\newblock


\bibitem[\protect\citeauthoryear{Callegari, Gazzarrini, Giordano, Pagano, and
  Pepe}{Callegari et~al\mbox{.}}{2011}]%
        {Callegari}
\bibfield{author}{\bibinfo{person}{C. Callegari}, \bibinfo{person}{L.
  Gazzarrini}, \bibinfo{person}{S. Giordano}, \bibinfo{person}{M. Pagano},
  {and} \bibinfo{person}{T. Pepe}.} \bibinfo{year}{2011}\natexlab{}.
\newblock \showarticletitle{A Novel PCA-Based Network Anomaly Detection}.
\newblock  (\bibinfo{year}{2011}), \bibinfo{pages}{1--5}.
\newblock


\bibitem[\protect\citeauthoryear{Fouladi, Emmons, Orbay, Wu, Wahby, and
  Winstein}{Fouladi et~al\mbox{.}}{2018}]%
        {Salsify}
\bibfield{author}{\bibinfo{person}{S. Fouladi}, \bibinfo{person}{J. Emmons},
  \bibinfo{person}{E. Orbay}, \bibinfo{person}{C. Wu}, \bibinfo{person}{R.S.
  Wahby}, {and} \bibinfo{person}{K. Winstein}.}
  \bibinfo{year}{2018}\natexlab{}.
\newblock \showarticletitle{Salsify: Low-Latency Network Video through Tighter
  Integration between a Video Codec and a Transport Protocol}.
\newblock \bibinfo{journal}{{\em 15th USENIX Symposium on Networked Systems
  Design and Implementation\/}} (\bibinfo{year}{2018}),
  \bibinfo{pages}{267--282}.
\newblock


\bibitem[\protect\citeauthoryear{Institute}{Institute}{2007}]%
        {eipr}
\bibfield{author}{\bibinfo{person}{Electrical Power~Research Institute}.}
  \bibinfo{year}{2007}\natexlab{}.
\newblock \showarticletitle{Advanced Metering Infrastructure (AMI)}.
\newblock  (\bibinfo{year}{2007}).
\newblock


\bibitem[\protect\citeauthoryear{Talpur}{Talpur}{2017}]%
        {AliM}
\bibfield{author}{\bibinfo{person}{A.M. Talpur}.}
  \bibinfo{year}{2017}\natexlab{}.
\newblock {\em \bibinfo{title}{{Congestion Detection in Software Defined
  Networks using Machine Learning}}}.
\newblock \bibinfo{thesistype}{Master's\ thesis}. \bibinfo{school}{University
  of Bremen}, \bibinfo{address}{Germany}.
\newblock


\end{thebibliography}

\end{document}